# Side-gated transport in FIB-fabricated multilayered graphene nanoribbons.**

*Jean-François Dayen, Ather Mahmood, Dmitry S. Golubev, Isabelle Roch-Jeune, Philippe Salles and Erik Dujardin**

Since the discovery of superconductivity in doped fullerene[1] and semiconducting behaviour of carbon nanotubes more than a decade ago, the electronic properties of graphitic materials have been extensively revisited.[2-4] Considering that the transport properties of these materials derive from those of flat graphite and its ultimately thin constituent, graphene, it has been proposed quite early on that controlling the morphology of flat graphitic domains could lead to a new graphene-based nanoelectronics.[5] Several ways of producing micrometer-scale domain of few nanometer high graphite film have been reported, including recurrent mechanical cleavage of pyrolytic graphite (HOPG) using adhesives,[6] micropatterning of HOPG followed direct[7, 8] or cantilever-mediated exfoliation[9], thermal decomposition of hydrocarbons[10] or epitaxial growth on metallic[11] or carbide surfaces.[12, 13] Eventually, single layer graphene itself has been obtained recently by either micromechanical exfoliation of HOPG [14, 15] or by thermal decomposition of SiC,[16] which marked the onset of a quickly expanding research activity on graphene.[17] In contrast to carbon nanotube devices, the performances of which are still crucially dependent on the contact area, thin graphite and graphene offer a large micrometric interface to be connected to macroscopic electrodes, while nanoscale electrodes and active features can be patterned directly in the same material.[17-19] If numerous examples of graphene patterning at the micron-scale have been demonstrated, the fabrication of nanometric motives in thin graphite and graphene still requires the development of dedicated protocols. In this Letter, we present the patterning, exfoliation and micromanipulation of thin graphitic discs which are subsequently connected and patterned into sub-100nm wide ribbons with a resist-free process using Focused Ion Beam (FIB) lithography and deposition. The electronic transport properties of the double side-gated nanoribbons are then investigated down to 40 K and interpreted with a simple model of 1D array of tunnelling junctions.

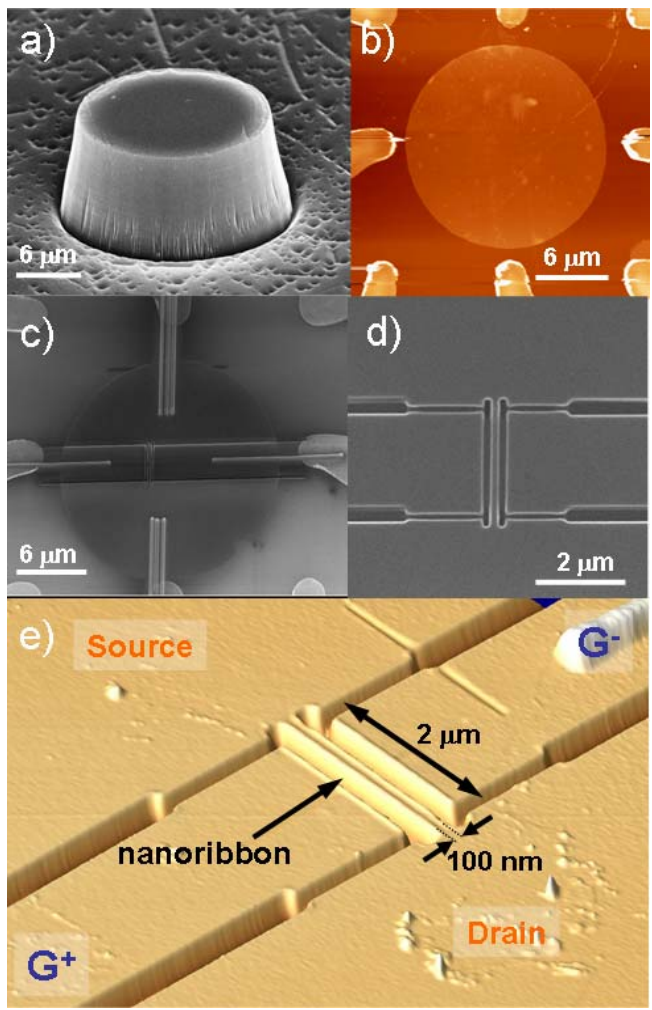

**Figure 1.** (a) SEM image of a 10 μm high and 20 μm diameter cylindrical HOPG pillar. (b) AFM image of an exfoliated, 12-nm thick graphite disk deposited at the centre of pre-patterned microelectrodes. (c) SEM image of a 30-layer graphite disc with metallic contacts and FIB-milled pattern; (d) SEM detail of the ribbon pattern with side gates which are separated by 150-nm wide trenches just after being cut; (e) 3D rendering of an AFM image of a side gated sub-100 nm wide graphite nano-ribbon.

[*] Dr. J.-F. Dayen, A. Mahmood, P. Salles, Dr. E. Dujardin
NanoSciences Group,
CEMES CNRS UPR 8011
BP 94347, 29 rue J. Marvig, 31055 Toulouse Cedex 4, France
Fax: (+) 33 5 62 25 79 99
E-mail: erik.dujardin@cemes.fr
Dr. D. S. Golubev
Forschungszentrum Karlsruhe, Institut für Nanotechnologie
76021 Karlsruhe, Germany
and
I.E. Tamm Department of Theoretical Physics
P.N. Lebedev Physics Institute, 119991 Moscow, Russia
Dr. I. Roch-Jeune
IEMN, UMR CNRS 8520
Av. H. Poincaré BP 60069, 59652 Villeneuve d'Ascq Cedex, France

[**] The authors thank G. Benassayag, D. Vuillaume and K.S. Novoselov for fruitful discussions. E.D. acknowledges the Agence National de la Recherche (Contracts ANR-JC05-46117 and ANR-06-NANO-019-02) and A.M. the High Education Commission of Pakistan.

The sample preparation consisted in four steps which are described in detail in the Experimental Section and illustrated in Figure 1. Briefly, 20 μm diameter and 10 μm high cylindrical micropillars are first patterned on a HOPG crystal (Fig. 1a).[7, 9] Secondly, thin disks are extracted from the core of the pillar and deposited at the centre of pre-patterned micro-electrodes with a tapered glass pipette mounted on a micromanipulator (Figure 1b). 12 to 8 nm thick disks were then obtained by in-situ exfoliation with the glass tip and the satisfactory structural quality was assessed by AFM imaging as well as Raman spectroscopy.[20] Next, 5 to 10 µm long, 30 nm thick and 100 to 200 nm wide metallic wires were directly deposited between the pre-patterned gold microelectrodes and the graphite disk using the FIB-assisted metal deposition of platinum, a better conductor than conventional tungsten, by decomposition of a gaseous precursor (Figure 1c). Indeed, while $Ga^+$ ion damage and contamination associated with FIB nanofabrication has prevented its use for making single walled nanotubes devices directly connected to metal electrodes, transport in multiwalled nanotubes,[3] and graphene[21] with FIB-fabricated contacts has been successfully reported. Although palladium is known to yield the best Schottky-barrier free contacts on nanotubes, gaseous precursors of palladium are not available for FIB deposition. These short-comings are not limiting in our case since contacts are large (typically one square micrometer) on a remote part of the graphite disk, which itself realizes the nanoelectrode contacts with the nanopatterned devices. Finally, a major asset of our approach is that milling graphite patterns of a few tens of nanometers in width can be performed directly in the connected graphitic domain without requiring any extra process nor using lithography resists which are known contaminants.[22] As shown in Figures 1c~e, a typical double side-gated transistor pattern consists in a 2 µm long ribbon, the nominal width of which was varied between of 500 and 150 nm. The two intact graphite hemi-circles act as source and drain electrodes. The two graphite side-gates are separated from the ribbon by a 100 nm wide gap. Importantly, the milling process generates an amorphous region around the cut features which reduces the actually conducting area and increases the gate-sample distance. The width of the amorphous zone could be estimated 65 ± 15 nm by SEM and AFM. Therefore, the conductive channel can be defined by an effective active width smaller than 350 nm for all investigated devices.

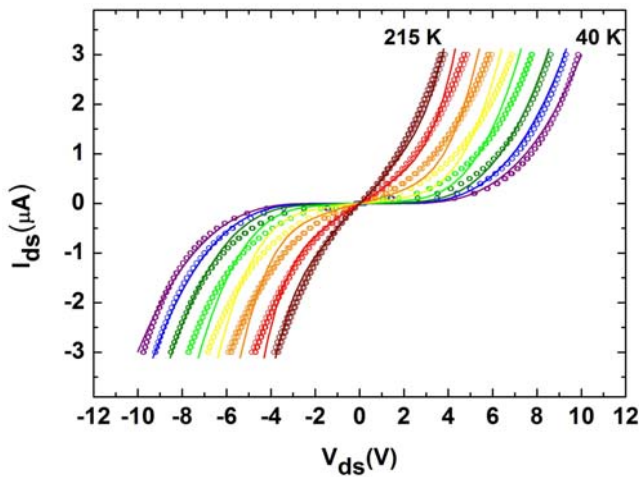

Figure 2. Temperature evolution of current-tension characteristic curves for a 50 nm × 2 µm, 30-monolayered ribbon at 215, 190, 165, 140, 115, 90, 67 and 40 K. Gate voltage is zero. Open circles are experimental data and continuous lines are derived from optimised fit using the 1D QD array model (see text).

The transport properties of double side-gated nanoribbons were measured from 300K to 40K, by applying DC source-drain current ramp under different gate DC voltages applied across the two side-gates, and by measuring the corresponding source-drain DC voltage drop on a Keithley Nanovoltmeter. In order to validate the entire fabrication process, non-patterned disks were contacted as described. These control samples had ohmic contact resistances comprised between 200 Ω and 800 Ω at 40 K, showing negligible temperature variation between 300 and 40 K and no bottom-gate dependency thus indicating the good quality of the FIB-deposited Pt contacts (Supporting Information, Figures S1 and S2). The channel resistance of the nanoribbons varies from sample to sample and ranges from 40 to 700 kΩ at 300 K. Noteworthy, ribbons with active width larger than 300 nm showed an ohmic behaviour without any side-gate effect ranging from a few kΩ at room temperature and reaching typically 100 kΩ at 40 K (Supporting information, Figure S3 and S4). Such observations validate our approach in which metallic contacts are made with the microscopic graphitic pads far from the sensitive ribbons while the two graphite semi-discs themselves are acting as the terminal electrode. In this way, not only do we avoid Schottky contacts on the nanoribbons but potential damages during Pt deposition are kept away from the sensitive area. Conversely, ribbons with a nominal width smaller than 150 nm showed a marked amorphization inside the channel and had resistances greater than 1 MΩ. Their strong blockade behaviour without measurable gate effect is not presented in this Letter. In practice, only ribbons that showed a marked gate voltage dependence in their I-V characteristics, which corresponded to nominal width comprised between 250 and 200 nm, or an effective width between 100 and 50 nm, were considered. Their resistance at room temperature was typically a few hundreds of kΩ reaching 10 to 30 MΩ at 40 K (Supporting information, Figure S5)

The temperature evolution of the $I_{ds}(V_{ds})$ characteristic for a typical 2 µm-long nanoribbon with a 50 nm active width is displayed in open circles on Figure 2. While the $I_{ds}(V_{ds})$ curve is close to linear at room temperature, a non-linear behaviour develops as the temperature decreases with a gap around zero bias increasing up to 10 V at 30 K. For slightly larger active width (e.g. 100 nm), the low temperature gap was observed to be significantly smaller (e.g. 6 V) and the characteristics became linear well below room temperature. The dependence of this non-linearity on side-gate voltage, $V_G$, was studied with $V_G$ ranging from -30V to +30V but only data for $V_G \geq 0$ are presented since the data were fully symmetrical. To illustrate the general observations, the data of a 50-nm wide ribbon measured at 40 K and 190 K are reported as open circles on Figure 3a and 3b respectively. For both temperatures, a symmetrical decrease of the gap amplitude with increasing $V_G$ is observed. Interestingly, fine structures shaped as additive steps are occasionally observed on some I(V) curves as is the case for $V_G = 10V$ and, to a lesser extent for $V_G = 15V$, on Figure 3a. Quite clearly, with the recorded I(V) characteristics, the ribbon device can not be described as a conventional field effect transistor, which could be expected from a graphitic ribbon, nor as a Schottky barrier transistor since no saturation is observed and such effects are not observed in the control experiments either. Instead, the non-linearity and the gate effect could be accounted for by a series of tunnelling barriers appearing in sufficiently narrow ribbons.

In order to explore further this interpretation, a model is proposed based on Coulomb blockade in a 1D array of N tunnel junctions, which gives a simple and intuitive picture of the observed electronic transport.[23] First we define the "non-interacting" current $I_0(V)$ through the tunnel junction separating two graphene quantum dots. Assuming a linear dispersion in graphene, we get:

$$I_0(V) = \frac{A}{k_B^3} \int dE |E - eV - \mu||E - \mu| \left( \frac{1}{1+\exp\left[\frac{E-eV}{k_BT}\right]} - \frac{1}{1+\exp\left[\frac{E}{k_BT}\right]} \right) \quad (1)$$

where μ is the chemical potential of the quantum dots and the pre-factor A is related to the junction transparency. In real metallic materials, μ depends linearly on the gate voltage, $\mu = e.C_G.V_G/C$, where $C_G$ is the gate capacitance and C is the total capacitance of a quantum dot. However, since graphene is a semi-metal, this dependence is not necessarily linear. If such a junction is embedded in a 1D array of N high ohmic junctions with the same resistance and Coulomb interaction is switched on, then the I(V) curve is modified as follows:[24]

$$I(V) = \frac{I_0(V/N - V_C)}{1-\exp\left[\frac{e(V_C - V/N)}{k_BT}\right]} + \frac{I_0(V/N + V_C)}{1-\exp\left[\frac{e(V_C + V/N)}{k_BT}\right]} \quad (2)$$

Here V/N is the voltage drop on one of the junctions, and $V_C = e/C$ is Coulomb gap. Experimental I(V) data as a function of temperature and gate voltage were compared to equation (2), where the adjusting parameters where A, N, $V_C$ and, when the gate voltage was considered, μ. Continuous lines in Figures 2 and 3 result from the best consistent fits of all data, which could be obtained quite univocally. In particular, assuming the Coulomb gap energy constant, its best value was found to be e.$V_C$ = 30 meV in order to match the apparent gap at both 40 and 190 K and for the whole range of gate voltage. It was then realized that keeping either N or A independent of the temperature could not account for the entire datasets whereas allowing these parameters to vary with the temperature yielded a very good overall agreement (Supporting Information, Figures S6-S8). The presented best fit where thus obtained with N decreased roughly linearly from 60 ± 2 down to 40 ± 2 as temperature increased from 40 to 210 K. Concomitantly, A increased exponentially from 4.5×10$^{-15}$ ± 0.5×10$^{-15}$ A.K$^{-3}$ to 23×10$^{-15}$ ± 2×10$^{-15}$ A.K$^{-3}$ over the same temperature range. For a comparison, a slightly larger ribbon, with an effective width of about 90 nm, the best fit yielded a Coulomb gap energy of 20 meV and a number of junction comprised between 10 and 5.

Considering that the ribbon is 2 μm long, the best fit value of N = 60 a low temperature indicates that the maximal size of the quantum dots is about 30 nm, which is consistent with the optimal value of gap energy value (30 meV) that corresponds to a typical quantum dot size of 10 to 30 nm depending on the assumed dielectric constant of the ribbon. Furthermore, as the temperature increases, the number of junctions decreases which means that thermal activation renders some of the tunnel junctions transparent thus changing the observed subset of junction within the 1D array. Obviously, this is accompanied by a variation of the average ribbon transparency since the increase of A can be correlated to a decrease in the overall tunnelling resistance.

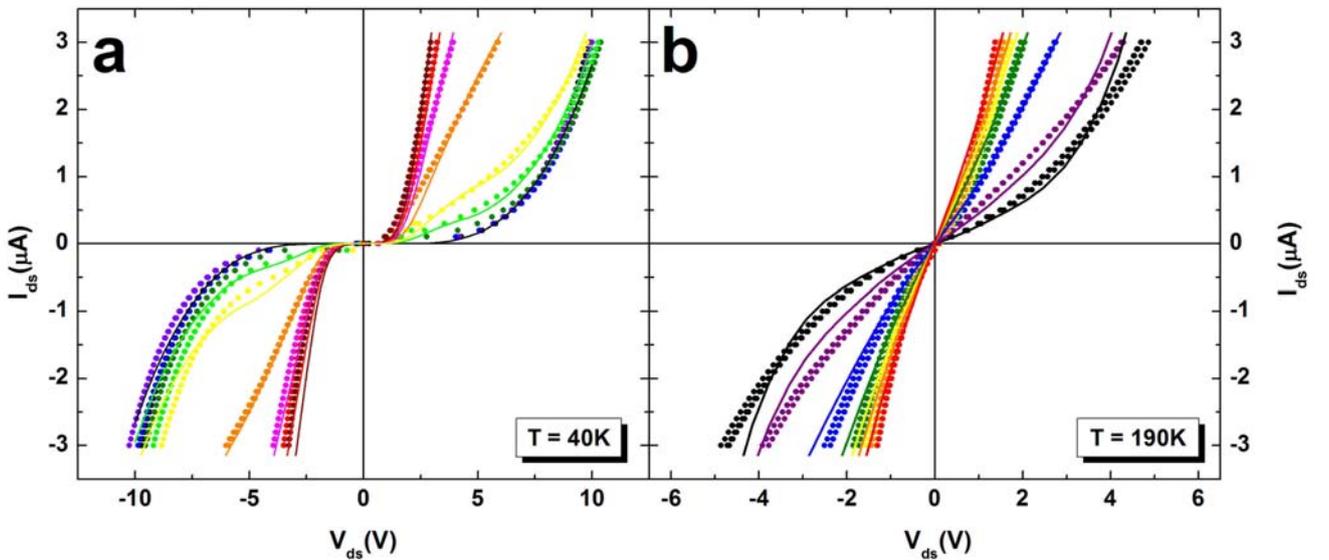

Figure 3. Evolution of the I-V characteristics of a 50 nm × 2 μm, 30-monolayered graphitic ribbon at (a) T = 40 K and (b) T = 190 K. Applied side-gate voltages are (a) 0; 2; 4; 6; 8; 10; 15; 20; 25; 30 V and (b) 0; 5; 10; 15; 20; 25; 30 V displayed in a black-to-red rainbow colour scheme. Open circles are experimental data and continuous lines are the best fits using the 1D QD array model with the following parameters : (a) N = 60; A = 4.5×10$^{-15}$ A.K$^{-3}$; $E_C$ = 30 meV and (b) N = 40; A = 18×10$^{-15}$ A.K$^{-3}$; $E_C$ = 30 meV (see text for details).

The nature of the tunnelling barriers in the ribbons is not clear although several suggestions can be made. In spite of our precautions, one can not exclude that a number of highly resistive defects acting as tunnelling barriers along the ribbon have been generated by the nanofabrication steps used in this study. For example native or ICP-RIE induced defects from the HOPG crystal, irregularities in the amorphous edges cut by FIB or even intercalated $Ga^{3+}$ ions. Moreover, inherent and random defects could be sufficient to account for such barriers.[25, 26] Alternatively, a random potential created by the underneath layers has been invoked in multiwalled carbon nanotubes devices,[27-29] and the corrugation of the graphene layers deposited on the $SiO_2$ substrate or even suspended[30] could be another source of tunnel junctions. Interestingly, conductance gaps in sub-100 nm wide monolayer graphene nanoribbons have been observed and interpreted either in terms of ribbon geometry and crystal structure[19] or by the presence of Coulomb gaps due to the existence of internal tunnelling junctions.[31] This last example and the present study demonstrate that reproducible nanofabrication process leads to well defined transport properties of laterally confined graphitic materials although the understanding of the intimate mechanisms still requires further experimental and theoretical investigations. Nevertheless, our method opens the way to actual transistor-like devices as illustrated in Figure 4, where the current across the nanoribbon is modulated by a factor 30 at 40 K upon switching the side-gate voltage. This characteristic is to be compared to the ON-OFF current ratio at room temperature of about 2 reported recently for the top-and-bottom gate[32] or the bottom gate[33] graphene field effect transistors.

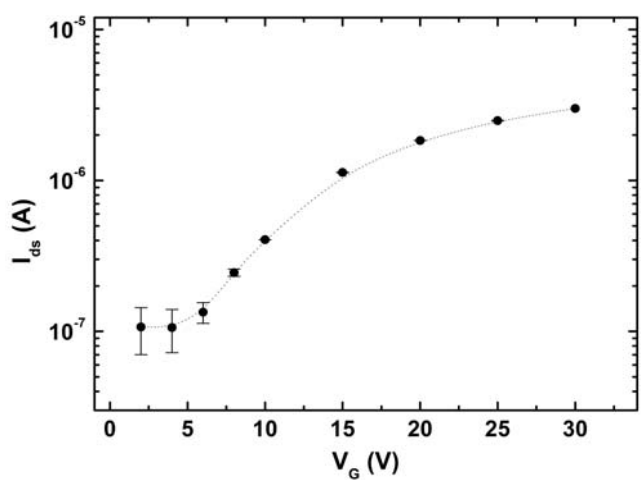

**Figure 4.** Transistor-like behaviour of the side-gated 50 nm × 2 µm, 30-monolayered nanoribbon. T = 40 K and $V_{ds}$ = 3 V.

In this Letter, we have fully detailed a resist-less method of fabricating and addressing nanoscale multilayer graphene devices and used the side-gated transistor device as a demonstrator. Cryotransport has been measured on sub-100 nm wide nanoribbons which revealed non-linear I(V) behaviour, the gap of which is controlled by the side-gate voltage. The experimental data have been interpreted as the manifestation of Coulomb blockade into a linear array of N tunnel junctions between graphene islands, by developing a simple and intuitive model. This framework puts under light the high sensitivity of such nanostructures to defects. Our study on thin graphitic nanoribbons is a first step in the understanding the electronic properties of laterally confined graphene samples which will be pursued on graphene monolayer in order to explore the potentials of graphene-based nanoelectronics.

**Experimental Section**

*Microfabrication of graphite micropillars.* A 700 nm layer of hydrogen silsesquioxane (HSQ, Dow Corning Fox-16) is spin-coated onto HOPG and 20-µm diameter circular patterns are defined by e-beam exposure (dose : 3500 µC/cm², Vistec Leica machine EBPG 5000 +) followed by development using AZ400 developer (Clariant). Densification of the resist and etching of the HOPG substrate are then performed simultaneously in an oxygen plasma using ICP-RIE machine (Plasmalab 100 - Oxford Instruments). Typical experimental conditions were: oxygen flow rate 20 sccm, pressure 6 mTorr, RIE power 125 W, ICP power 1500 W and DC bias -230 V for 15 minutes, leading to 10 µm high and 20 µm diameter cylindrical pillars. Finally, the HSQ film was removed in a HF solution.

*Electrical contacting of multilayered graphene disks.* Sub-20 nm thick disks are extracted from the core of the micropillar and deposited at the centre of pre-patterned micro-electrodes. This is performed with a tapered glass pipette having a ca. 1 µm tip apex radius mounted on a micromanipulator. Initially, the 3 to 5 µm uppermost portion of the pillar, which has undergone some ion irradiation during the ICP-RIE step, is removed and a ca. 1 µm-thick disk is exfoliated and placed in between the micro-electrodes. The graphite disk is then thinned down in-situ by successive peel-off of the upper layers. With sharp glass tips, disks with thicknesses ranging from 12 down to 8 nm could be routinely obtained. The electrical contacting is achieved by the direct deposition of platinum wires using a dual FIB-FEGSEM system (Zeiss-Orsay Physics 1540XB) assisted with a gas injection system that delivers the organometallic precursor $((CH_3)_3\ (CH_3C_5H_4)Pt)$. 5 to 10 µm long, 30 nm thick and 100 to 200 nm wide platinum wires were deposited between the pre-patterned gold microelectrodes and the graphite disk using 30 kV $Ga^{3+}$ beam. Typical irradiation dose were 2.5 to 4 nC.µm$^{-2}$ with an ion current stabilized around 65 pA.

*Tailoring the graphene sheets into devices.* After electrical connection, arbitrary patterns can be milled by FIB into the disks with a current of 5 to 9 pA at a low ion dose of 0.3 to 0.5 nC.µm$^{-2}$. Here, the tailoring of double side-gated transistor devices is reported where two half-disk graphitic electrodes are linked by a 2 µm long ribbon, the nominal width of which is chosen between 500 and 150 nm. Moreover, two graphite side gates are produced simultaneously.

[1] K. Tanigaki, T. W. Ebbesen, S. Saito, J. Mizuki, J. S. Tsai, Y. Kubo, S. Kuroshima, *Nature* **1991**, *352*, 222-223.
[2] R. Saito, G. Dresselhaus, M. S. Dresselhaus, *Physical Properties of Carbon Nanotubes*, Imperial College Press, London **1998**.
[3] T. W. Ebbesen, H. J. Lezec, H. Hiura, J. W. Bennett, H. F. Ghaemi, T. Thio, *Nature* **1996**, *382*, 54-56.

[4] J. C. Charlier, X. Blase, S. Roche, *Rev. Mod. Phys.* **2007**, *79*, 677-732.
[5] T. W. Ebbesen, H. Hiura, *Advanced Materials* **1995**, *7*, 582-586.
[6] H. Fernandez-Moran, *J. Appl. Phys.* **1960**, *31*, 1840.
[7] X. K. Lu, H. Huang, N. Nemchuk, R. S. Ruoff, *Appl. Phys. Lett.* **1999**, *75*, 193-195.
[8] Y. B. Zhang, J. P. Small, M. E. S. Amori, P. Kim, *Phys. Rev. Lett.* **2005**, *94*, 176803.
[9] Y. B. Zhang, J. P. Small, W. V. Pontius, P. Kim, *Appl. Phys. Lett.* **2005**, *86*, 073104.
[10] A. Krishnan, E. Dujardin, M. M. J. Treacy, J. Hugdahl, S. Lynum, T. W. Ebbesen, *Nature* **1997**, *388*, 451-454.
[11] T. A. Land, T. Michely, R. J. Behm, J. C. Hemminger, G. Comsa, *Surf. Sci.* **1992**, *264*, 261-270.
[12] H. Itoh, T. Ichinose, C. Oshima, T. Ichinokawa, T. Aizawa, *Surf. Sci.* **1991**, *254*, L437-L442.
[13] I. Forbeaux, J. M. Themlin, A. Charrier, F. Thibaudau, J. M. Debever, *Appl. Surf. Sci.* **2000**, *162*, 406-412.
[14] K. S. Novoselov, A. K. Geim, S. V. Morozov, D. Jiang, Y. Zhang, S. V. Dubonos, I. V. Grigorieva, A. A. Firsov, *Science* **2004**, *306*, 666-669.
[15] K. S. Novoselov, D. Jiang, F. Schedin, T. J. Booth, V. V. Khotkevich, S. V. Morozov, A. K. Geim, *Proc. Natl. Acad. Sci. U. S. A.* **2005**, *102*, 10451-10453.
[16] C. Berger, Z. M. Song, T. B. Li, X. B. Li, A. Y. Ogbazghi, R. Feng, Z. T. Dai, A. N. Marchenkov, E. H. Conrad, P. N. First, W. A. de Heer, *J. Phys. Chem. B* **2004**, *108*, 19912-19916.
[17] A. K. Geim, K. S. Novoselov, *Nature Materials* **2007**, *6*, 183-191.
[18] E. Dujardin, T. Thio, H. Lezec, T. W. Ebbesen, *Appl. Phys. Lett.* **2001**, *79*, 2474-2476.
[19] M. Y. Han, B. Ozyilmaz, Y. B. Zhang, P. Kim, *Phys. Rev. Lett.* **2007**, *98*, 206805.
[20] C. Faugeras, A. Nerrière, M. Potemski, A. Mahmood, E. Dujardin, C. Berger, W. A. de Heer, *Appl. Phys. Lett.* **2008**, *92*, 011914.
[21] A. Shailos, W. Nativel, A. Kasumov, C. Collet, M. Ferrier, S. Gueron, R. Deblock, H. Bouchiat, *EPL* **2007**, *79*, 57008.
[22] M. Ishigami, J. H. Chen, W. G. Cullen, M. S. Fuhrer, E. D. Williams, *Nano Lett.* **2007**, *7*, 1643-1648.
[23] P. Delsing, in *Single Charge Tunneling*, Vol. 294 (Eds: H. Grabert, M. H. Devoret), Plenum, New York **1992**, 249-274.
[24] D. V. Averin, K. K. Likharev, in *Mesoscopic Phenomena in Solids*, (Eds: B. L. Altshuler, P. A. Lee, R. A. Webb), Elsevier Science Publisher B. V., **1991**, 173-271.
[25] Y. V. Nazarov, *Phys. Rev. Lett.* **1999**, *82*, 1245-1248.
[26] S. Farhangfar, R. S. Poikolainen, J. P. Pekola, D. S. Golubev, A. D. Zaikin, *Phys. Rev. B* **2001**, *63*, 075309.
[27] R. Egger, A. O. Gogolin, *Chem. Phys.* **2002**, *281*, 447-454.
[28] P. L. McEuen, M. Bockrath, D. H. Cobden, Y. G. Yoon, S. G. Louie, *Phys. Rev. Lett.* **1999**, *83*, 5098-5101.
[29] M. J. Biercuk, N. Mason, J. M. Chow, C. M. Marcus, *Nano Lett.* **2004**, *4*, 2499-2502.
[30] J. C. Meyer, A. K. Geim, M. I. Katsnelson, K. S. Novoselov, T. J. Booth, S. Roth, *Nature* **2007**, *446*, 60-63.
[31] F. Sols, F. Guinea, A. H. C. Neto, *Phys. Rev. Lett.* **2007**, *99*, 166803.
[32] M. C. Lemme, T. J. Echtermeyer, M. Baus, H. Kurz, *IEEE Electron Device Lett.* **2007**, *28*, 282-284.
[33] X. Liang, Z. Fu, S. Y. Chou, *Nano Lett.* **2007**, *7*, 3840-3844.

# Side-gated transport in FIB-fabricated multilayered graphene nanoribbons.

*Jean-François Dayen, Ather Mahmood, Dmitry S. Golubev, Isabelle Roch-Jeune, Philippe Salles and Erik Dujardin\**

## Supporting Information

**Figure S1:** I(V) curves for T varying between 300 and 42 K measured on a Pt-contacted, non-patterned graphite microdisc of similar characteristics as the ones used to fabricate side-gated nanoribbons. Inset: SEM picture of a Pt-contacted, non-patterned graphite microdisc.

**Figure S2:** Thermal variation of the 2-terminal resistance of a Pt-contacted non-patterned graphite microdisc in the 300 to 40 K range.

**Figure S3:** I(V) curves at 37, 88 and 135 K measured for gate voltage varying between -20 V and +20 V on a Pt-contacted graphene nanoribbon with nominal width of 350 nm. Inset: SEM picture of a side-gated device composed of a 300-nm wide nanoribbon.

**Figure S4:** Thermal variation of the 2-terminal resistance of a side-gated nanoribbon device with width of 350 nm in the 300 to 40 K range.

**Figure S5:** Thermal variation of the 2-terminal resistance of a side-gated nanoribbon device with width of 100 nm in the 220 to 40 K range.

**Figure S6:** Temperature dependence of the number of tunnel junction determined by fitting the experimental transport data of a 50 nm × 2 µm, 30-monolayered graphitic ribbon with a model consisting of a 1D array of tunnel junctions. The red dotted line is a linear fit.

**Figure S7:** Temperature dependence of the current pre-factor A determined by fitting the experimental transport data of a 50 nm × 2 µm, 30-monolayered graphitic ribbon with a model consisting of a 1D array of tunnel junctions. A can be related to the apparent tunnel transparency. The red dotted line is an exponential fit.

**Figure S8:** Side-gate voltage dependence of the chemical potential determined by fitting the 40 K and 190 K experimental transport data of a 50 nm × 2 µm, 30-monolayered graphitic ribbon with a model consisting of a 1D array of tunnel junctions. The red line is a linear fit to the low temperature data.

[*]   Dr. J.-F. Dayen, A. Mahmood, P. Salles, Dr. E. Dujardin
      NanoSciences Group,
      CEMES CNRS UPR 8011
      BP 94347, 29 rue J. Marvig, 31055 Toulouse Cedex 4, France
      Fax: (+) 33 5 62 25 79 99
      E-mail: erik.dujardin@cemes.fr
      Dr. D. S. Golubev
      Forschungszentrum Karlsruhe, Institut für Nanotechnologie
      76021 Karlsruhe, Germany
      and
      I.E. Tamm Department of Theoretical Physics
      P.N. Lebedev Physics Institute, 119991 Moscow, Russia
      Dr. I. Roch-Jeune
      IEMN, UMR CNRS 8520
      Av. H. Poincaré BP 60069, 59652 Villeneuve d'Ascq Cedex,
      France

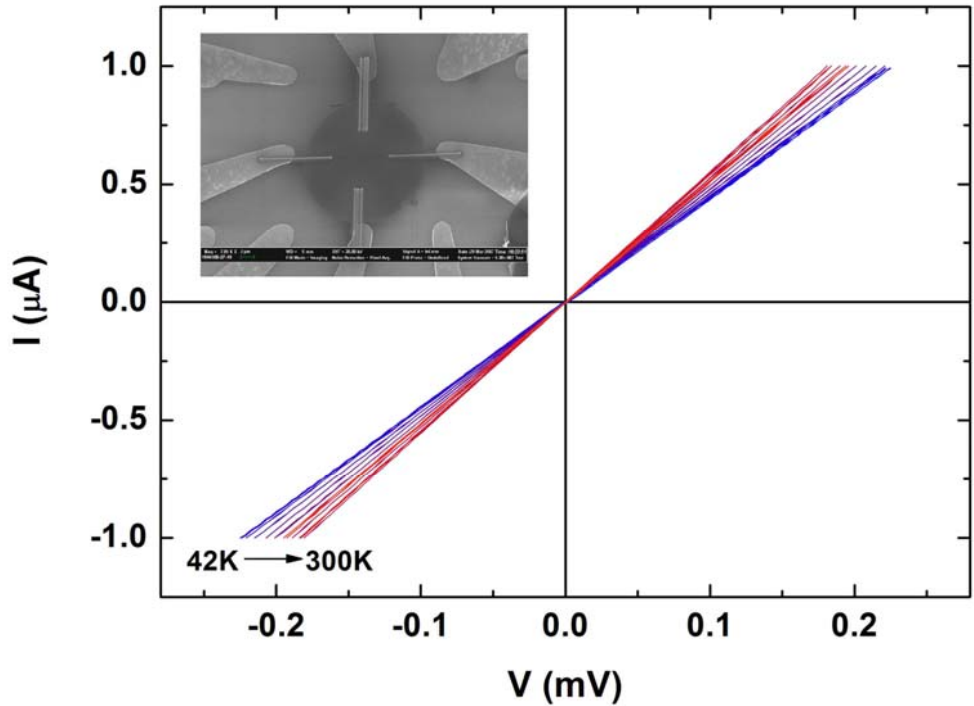

**Figure S1:** I(V) curves for T varying between 300 and 42 K measured on a Pt-contacted, non-patterned graphite microdisc of similar characteristics as the ones used to fabricate side-gated nanoribbons. Inset: SEM picture of a Pt-contacted, non-patterned graphite microdisc.

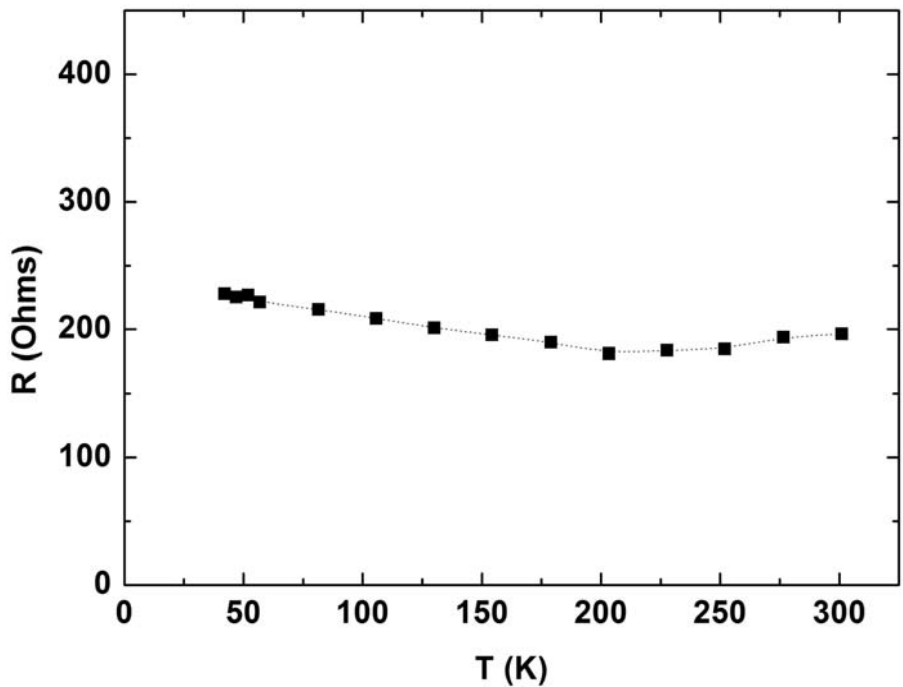

**Figure S2:** Thermal variation of the 2-terminal resistance of a Pt-contacted non-patterned graphite microdisc in the 300 to 40 K range.

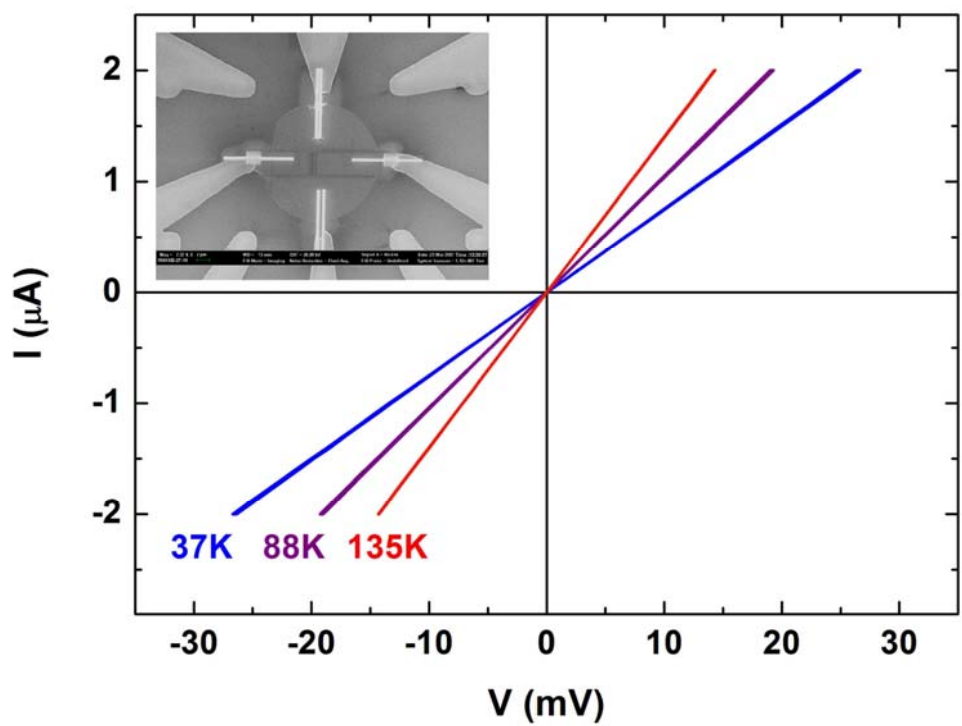

**Figure S3:** I(V) curves at 37, 88 and 135 K measured for gate voltage varying between -20 V and +20 V on a Pt-contacted graphene nanoribbon with nominal width of 300 nm. Inset: SEM picture of a side-gated device composed of a 350-nm wide nanoribbon.

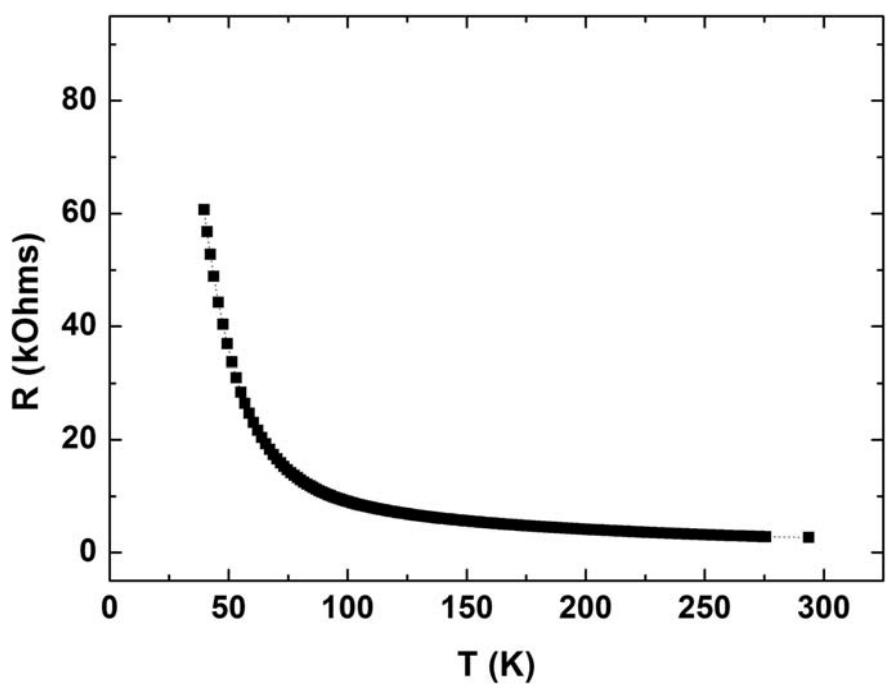

**Figure S4:** Thermal variation of the 2-terminal resistance of a side-gated nanoribbon device with width of 350 nm in the 300 to 40 K range.

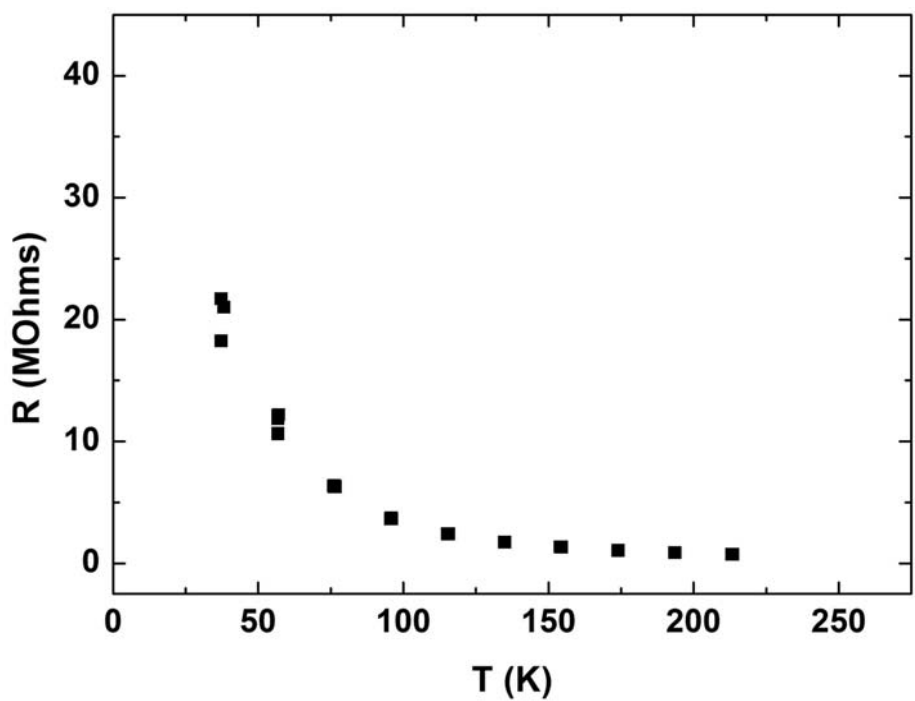

**Figure S5:** Thermal variation of the 2-terminal resistance of a side-gated nanoribbon device with width of 100 nm in the 220 to 40 K range.

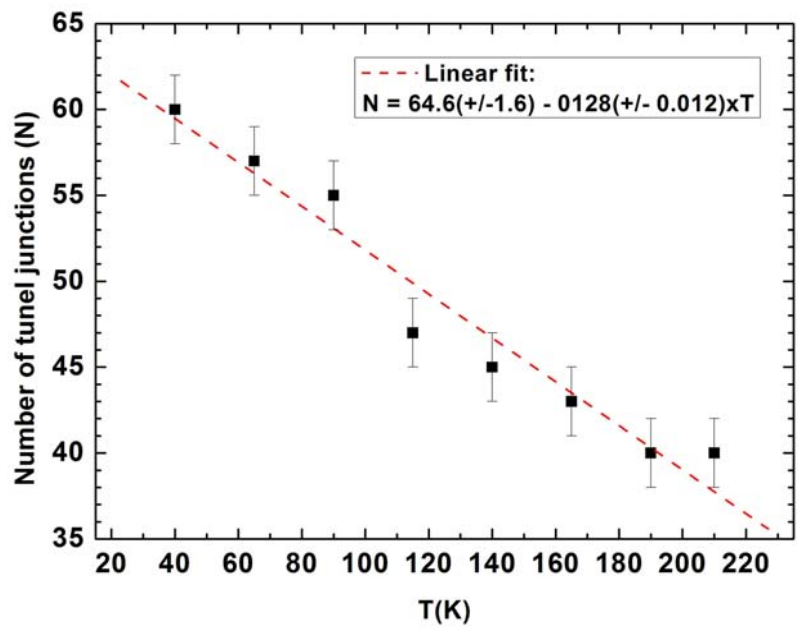

**Figure S6:** Temperature dependence of the number of tunnel junction determined by fitting the experimental transport data of a 50 nm × 2 μm, 30-monolayered graphitic ribbon with a model consisting of a 1D array of tunnel junctions. The red dotted line is a linear fit.

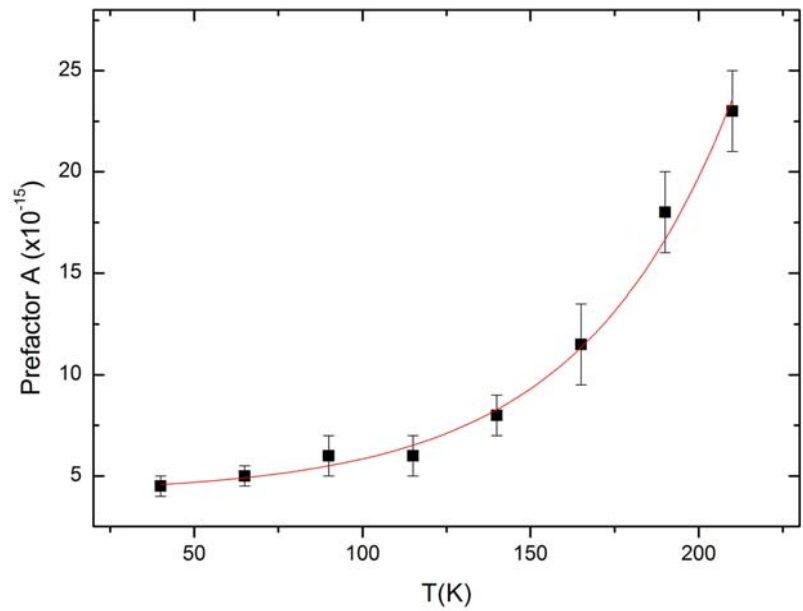

**Figure S7:** Temperature dependence of the current pre-factor A determined by fitting the experimental transport data of a 50 nm × 2 μm, 30-monolayered graphitic ribbon with a model consisting of a 1D array of tunnel junctions. A can be related to the apparent tunnel transparency. The red dotted line is an exponential fit.

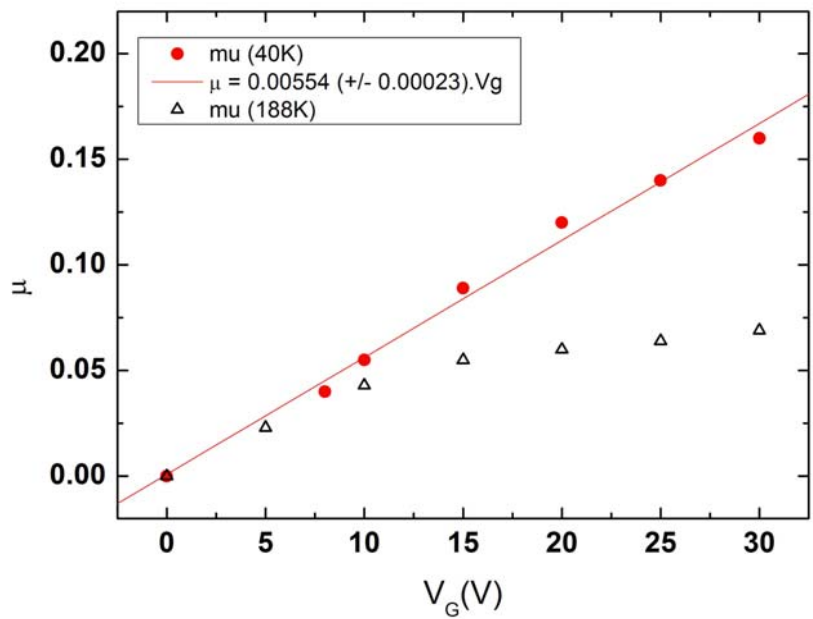

**Figure S8:** Side-gate voltage dependence of the chemical potential determined by fitting the 40 K and 190 K experimental transport data of a 50 nm × 2 μm, 30-monolayered graphitic ribbon with a model consisting of a 1D array of tunnel junctions. The red line is a linear fit to the low temperature data.